# A PARAMETRIC STUDY ON WINDOW-TO-FLOOR RATIO OF THREE WINDOW TYPES USING DYNAMIC SIMULATION

**Ana Rita Amaral**[1]\*, **Eugénio Rodrigues**[1,2],
**Adélio Rodrigues Gaspar**[1] and **Álvaro Gomes**[2,3]

1: ADAI, LAETA & Department of Mechanical Engineering
Faculty of Sciences and Technology
University of Coimbra
Rua Luís Reis Santos, Pólo II, 3030-788 Coimbra, Portugal
\* e-mail: aritamaral@hotmail.com, web: http://www.uc.pt/en/efs/research/geraplano

2: INESC Coimbra – Institute for Systems Engineering and Computers at Coimbra
Rua Antero de Quental 199, 3000-033 Coimbra, Portugal

3: Department of Electrical and Computer Engineering,
Faculty of Sciences and Technology,
University of Coimbra
Pólo II, 3030-290 Coimbra, Portugal

**Abstract** *The windows can be responsible for unnecessary energy consumption in a building, if incorrectly designed, shadowed or oriented. Considering an annual thermal comfort assessment of a space, if windows are over-dimensioned, they can contribute to the increase of the heating needs due to heat losses, and also to the increase of cooling needs due to over-exposure to solar radiation. When under-dimensioned, the same space may benefit from reduced heat losses through the glazing surface but does not benefit from solar radiation gains. Therefore, it is important to find the optimum design that minimizes both the heating and cooling needs. This paper presents a parametric study of window type (single, double and triple glazing), orientation and opening size, located in the city of Coimbra, Portugal. An annual and a seasonal assessment were done, in order to obtain the set of optimum values around 360º orientation.*

## 1. INTRODUCTION

Design decisions taken in early architectural design process phase may significantly contribute to the final building energy performance. From all building aspects, windows play an important role in thermal performance; an inappropriate area or type of glazing may increase negatively heat gains or losses, which can affect the thermal comfort of its occupants. Several studies have already focused in this subject, namely the use of dynamic simulation tools to predict the gains and losses through windows [1]. Hee et al. conducted an extensive literature research that congregates several aspects studied by numerous research groups and that may influence the global performance of a window, such as glazing type, fenestrations products and materials, spacers, frames, air or gas gap between glass layers in glazing systems, and so on [2]. Moreover, the use of triple glazing is not yet consensual. For instance, Tahmasebi et al. compared thermal performance of double and triple glazed windows and concluded that triple glazing contributes to the reduction of annual energy consumption in all quadrants and in different Window-to-Floor Ratios (WFR) [3]. On other hand, Gasparella et al. emphasize the fact that triple glazing windows show low solar transmittance, which in winter season can contribute to the reduction of solar gains that can overcome the reduction of thermal losses and increase energy needs [4].

From the literature survey, it stood out the fact that researchers mainly focused on advances of window technologies. However, the present study differs by proposing an approach that congregates several aspects, as glazing type, window size, and orientation, taking into consideration that geographical location and building systems provide specific conditions, potentially leading to different results. The main purpose is to assist practitioners, namely architects, to take conscious decisions in an early phase of design process, by providing data in terms of windows thermal performance. This assessment was made in an annual, orientation quadrant, and seasonal base.

## 2. METHODOLOGY

### 2.1. Reference room

The methodology used in the present study considers a reference room with fixed dimensions. Figure 1 illustrates the geometry of the reference room. In the widest wall, it is placed the opening to be analyzed. Starting with an opening width of 0.01m and a height of 2.00m, the opening will increase until it reaches 7.00m wide, thus having a WFR ranging from 0 to 0.614. Also, the opening orientation is analyzed in a two-degree step around the 360º, starting at 0º (North) and turning then to East.

The envelope constructive system is presented in Table 1 with the reference U-values of the Portuguese regulation [5] satisfied for the floor, roof, and exterior wall elements, knowing that Coimbra is integrated in the climatic zone I1 for winter and V2 for summer. The physical properties of each material layer are also listed. Three glazing physical properties were chosen to characterize Single (SGW), Double (DGW), and Triple Glazing Window (TGW) with 5.70, 2.60, and 1.00 U-values respectively.

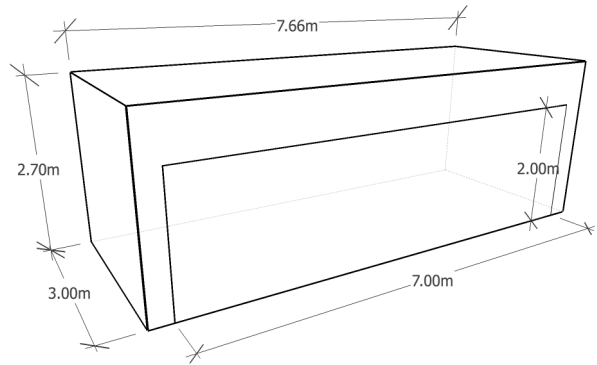

Figure 1: Thermal space geometric specifications for the parametric study.

Table 1: Reference U-values and materials physical properties

| Element | U-value | Type | SHGC | VT | Element | U-value | Type |
|---|---|---|---|---|---|---|---|
| Window | 5.70 | Single Glass | 0.66 | 0.70 | Ext. Wall | 0.43 | Double Brick Wall |
| | 2.60 | Double Glass | 0.63 | 0.56 | Floor | 0.45 | Ground Floor |
| | 1.00 | Triple Glass | 0.51 | 0.42 | Roof | 0.37 | Flat Roof |

SHGC – solar heat gain coefficient; VT – visible transmittance

## 2.2. Thermal comfort assessment

The thermal comfort assessment was calculated by summing the total degree-hours of thermal discomfort (TDH). The heating degree-hours (HDH) and cooling degree-hours (CDH) were determined by obtaining the difference between the operative indoor temperature in the reference room and adaptive thermal comfort limits for naturally ventilated spaces, according to the European Standard 15251:2007. The operative temperature in the reference room was obtained by using a dynamic simulation program (EnegyPlus 8.1.0). In order to achieve a global optimum window size, applicable to different building types, it is considered that other factors that may interfere in the results by an hypothetic number chosen for study may influence the energy consumption by heat gains or losses, are valueless. In this sense, the room was considered to have no occupation and no internal gains from equipment and lighting. However, infiltration was considered as a steady 0.4 Air Changes per Hour (ACH). As the room was modeled not to have any occupation, the openings were set close all time, thus not having any additional natural ventilation. The weather data was the one considered to Coimbra, Portugal, retrieved from the US Department of Energy website.

## 2.3. A parametric approach

For this parametric study were used two algorithms, EPSAP and FPOP, specifically adjusted to carry out this work [6–10]. EPSAP (Evolutionary Program for the Space Allocation Problem) consists in a hybrid evolutionary strategy approach enhanced with local search technique to allocate rooms on a floor plan [6–8]. FPOP (Floor plan

Performance Optimization Program) consists in a sequential variable optimization procedure, where different building geometry variables, such as the openings orientation, position and size, are changed, with the aim of minimizing the thermal penalties. These are obtained by the determination of the total degree-hours of discomfort [9,10]. The thermal performance of each solution is estimated using the dynamic simulation program EnergyPlus (v.8.1.0). FPOP has stored in its database constructive systems, with materials physical properties, which is in accordance with the Portuguese building regulation [11], as well as weather data and location information.

The room requirements were introduced in the EPSAP algorithm, which generated 180 reference rooms with an orientation difference of two degrees. After the reference rooms have been generated, the FPOP algorithm transformed the size of the window in exterior façade until it reached 7.00m wide and performed the dynamic simulation. This process was repeated for the three window types. In total, 76140 simulation runs were carried out.

## 3. RESULTS AND DISCUSSION

From the methodology described above, it was possible to obtain a global and a seasonal assessment for different orientations. In Table 2 are listed the intervals of optimum WFR for annual assessment, quadrant annual assessment, and seasonal assessment. It is observable that WFR varies greatly, e.g. DGW has an interval of optimum WFR between 0.126 and 0.387. This behavior is even stronger when the assessment is seasonal. In wintertime, the WFR tends to increase up to filling the available wall area, and in the case of summertime, as explained before, the opening could even be absent as it does not improve the space performance.

Table 2: WFR optimum interval per window type for Coimbra region.
* Summer does not have any thermal discomfort up to this WFR value.
Bold values represent the maximum WFR.

| Type | Annual | Orientation Quadrant | | Season | |
|---|---|---|---|---|---|
| SGW | 0.126 – 0.318 | North | 0.170 – 0.318 | Winter | 0.331 – **0.614** |
| | | East | 0.148 – 0.200 | Spring | 0.157 – 0.453 |
| | | South | 0.152 – 0.178 | Summer | 0.165* |
| | | West | 0.126 – 0.170 | Autumn | 0.187 – 0.348 |
| DGW | 0.126 – 0.387 | North | 0.178 – 0.387 | Winter | 0.296 – **0.614** |
| | | East | 0.152 – 0.231 | Spring | 0.139 – 0.435 |
| | | South | 0.152 – 0.183 | Summer | 0.196* |
| | | West | 0.126 – 0.178 | Autumn | 0.183 – **0.614** |
| TGW | 0.148 – 0.409 | North | 0.222 – 0.409 | Winter | 0.318 – **0.614** |
| | | East | 0.178 – 0.265 | Spring | 0.161 – 0.413 |
| | | South | 0.187 – 0.231 | Summer | 0.240* |
| | | West | 0.148 – 0.222 | Autumn | 0.213 – **0.614** |

When windows performance is split by season, it is observable that in winter and autumn the WFR variation is even greater than in the annual performance. In the springtime and summertime, the behavior is different. In the case of TGW, in spring, the performance is similar in any orientation. Yet, in summer, any window type is negligible for the thermal comfort, however WFR may go up to 0.165, 0.196, and 0.240, respectively SGW, DGW, and TGW, before the space gets worst performance due to overheating.

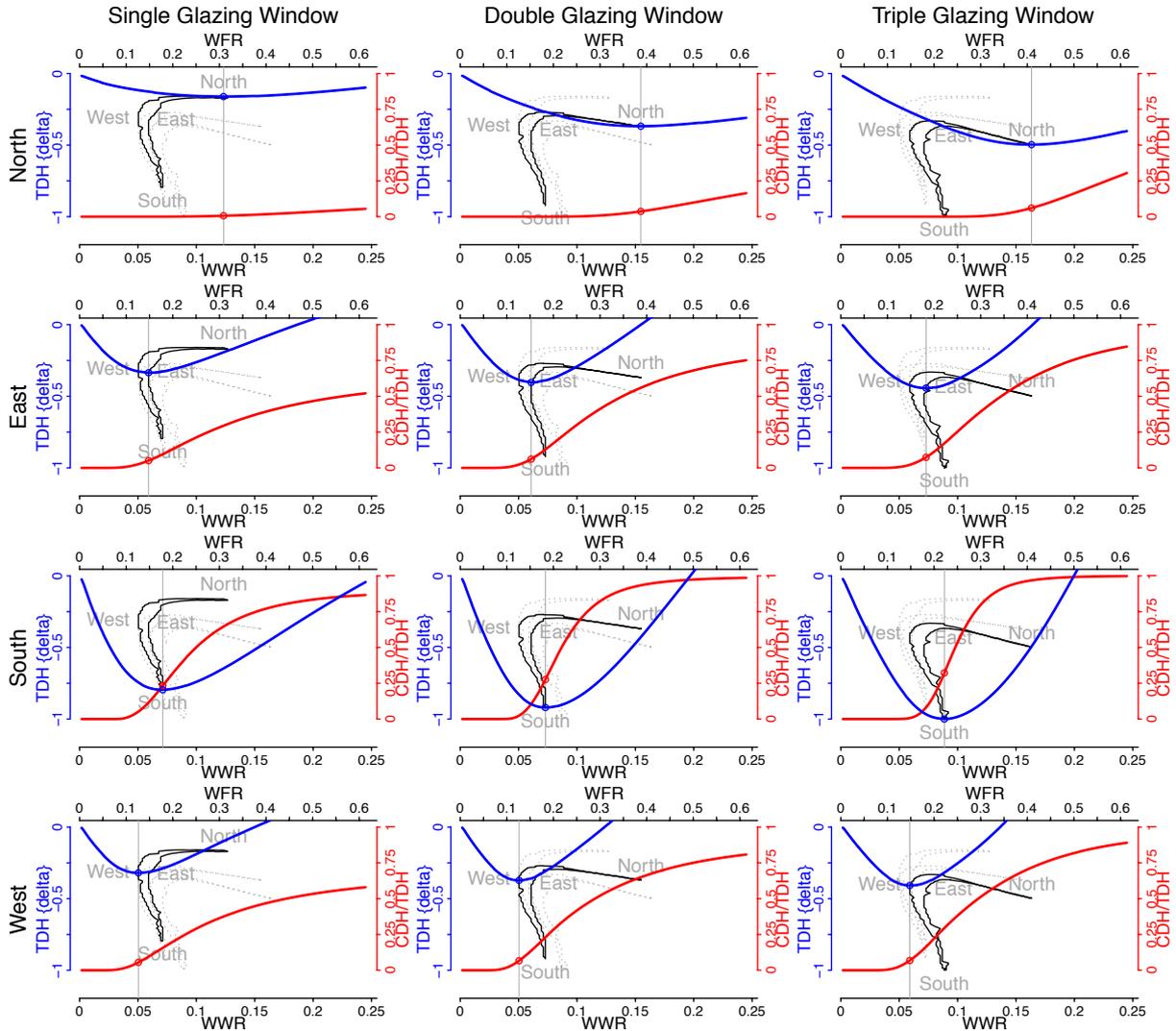

Figure 2: Annual assessment of the window size. The blue line represents the relative space thermal comfort performance (TDH), the red line the ratio of CDH per TDH, the thin black line represents the optimum window sizes for all orientations, and the grey dotted lines represent the other window type optimum values.

Figure 2 depicts the optimum performance points for the 360° orientations for annual assessment. The left-axis represents the space relative thermal performance from a space without any opening to the best opening size of all orientations. The top and bottom axis represent the WFR and Window-to-Wall Ratio (WWR), respectively. The right-axis is the ratio between CDH and TDH. The thin black line represents all the minimum values for all orientations, and red line the relative amount of overheating. As shown in Figure 2, it is noticeable that TGW has the best performance. However, the advantage of this kind of window is mainly observable when the opening is facing the North quadrant. For all window types, an unexpected behavior is evident in this quadrant. They have a higher WFR than in any other orientation. From a detailed analysis of the heat gains and losses it is concluded that this is due to window sky diffuse radiation gain being higher than the additional losses (relatively to opaque wall surface) from heat transfer to the exterior by the window surface. In the case of the DGW and TGW, the performance is better than surrounding orientations. For instance, in the case of TGW, North orientation is even better than East and West opening orientations. The reason for this behavior may result from the fast increase of CDH in East and West orientations due to solar beam radiation gain. As expected, the ratio of CDH per TDH is higher in this window type (red line). Still, this ratio is far from representing half of TDH when optimum window size found in any quadrant, thus denoting that the heat losses through the window surface are higher than the solar heat gains.

## 4. CONCLUSIONS

Despite not considering other performance requirements, such as lighting conditions and visual comfort, from the mere thermal comfort assessment it is possible to understand how window types, size, and orientation can have significant differences. Also, it was possible to conclude that large glazing areas facing north are not particularly bad for the thermal performance, since gains through sky diffuse radiation compensate possible thermal losses. However, the benefits are almost none in the case of windows with low thermal resistance. In the case of seasonal assessment, it is evident that WFR vary greatly. From wintertime, where window size tends to be the largest possible, to summertime, where even the existence of the opening is dispensable, the space performance varies significantly and an ideal opening WFR is difficult to determine, as a single value may hide different thermal behaviors around the year. In a global way, it is known that different types of glazing perform differently in diverse orientations. However, through this approach, it is possible to obtain the minimum values of penalties for all orientations. This means that it is possible to consider the optimal window size in new buildings, according to orientation, location and buildings physical properties, and thus considering the best glazing type either in new or in existing buildings, in an early phase of design process. Moreover, this analysis can be beneficial when designing windows in buildings with seasonal occupation, whether in winter or in summer.

This tool allows a simple and intuitive assessment of building systems, namely openings, providing useful data that can help architects and decision makers to obtain the solution with less thermal penalties for each case. With the combination of the simulation tool, it is possible to extrapolate this study for all existing climate data, being helpful to the overall designer's community.

ACKNOWLEDGEMENTS

This work is framed under the *Energy for Sustainability Initiative* of the University of Coimbra (UC) and has been supported by the project *Automatic Generation of Architectural Floor Plans with Energy Optimization* - GerAPlanO - QREN 38922 Project (CENTRO-07-0402-FEDER-038922).